# The One-Dimensional Boltzmann Equation for the Heat Transport Induced by Ultra-Short Laser Pulses

Janina Marciak-Kozlowska[a] and Miroslaw Kozlowski[b,*]

[a] *Institute of Electron Technology, Al. Lotników 32/46, 02-668 Warsaw, Poland.*
[b] *Science Teacher College and Institute of Experimental Physics, Physics Department, Warsaw University, Hoza 69, 00-668 Warsaw, Poland, e-mail: mirkoz@fuw.edu.pl*

[*] *Corresponding author*

**Abstract**
In this paper the Boltzmann transport equation for thermal processes induced by ultra-short laser pulses is formulated and solved. For thermal process where the duration of the laser pulse (femtoseconds - attoseconds) is shorter than the relaxation time, the solution of Boltzmann equation differs from the solution of the Fourier equation. In this paper the formula for density current of the heat carrier is obtained. It is shown that for thin one-dimensional structures the current strongly depends on the scattering mechanism of the heat carriers.
Key words: Boltzmann equation; Ultra-short laser pulses; Nonhomogenous materials.

# 1. Introduction

Recently it has been shown that after optical excitation by femtosecond pulse establishment of an electron temperature by e-e scattering takes place on a few hundred femtosecond time scale in both bulk and nanostructured noble materials [1–4]. In noble metal clusters the electron thermalization time (relaxation time) is of the order of 200 fs [3,4]. This relaxation time is much larger than the duration of the now available femtosecond optical pulses offering the unique possibility of analyzing the properties of a thermal quasi-free electron gas [5]. In paper [5] using a two color femtosecond pump-probe laser technique the ultrafast energy exchanges of a nonequilibrium electron gas was investigated. When the duration of the laser pulse, 25 fs, in paper [5], is shorter than the relaxation time the parabolic Fourier equation cannot be used [6,7]. Instead, the new hyperbolic quantum heat transport equation is the valid equation [8]. The quantum heat transport equation is the wave damped equation for heat phenomena on the femtosecond scale.

Wave is an organized propagating imbalance [10]. Some phenomena seem to be clearly diffusive, with no wave-like implications, heat for instance. That was consistent with experiments at the late century, but not any longer. As far back as the 1960s ballistic (wave-like) heat pulses were observed at low temperatures [10]. The idea was that heat is just the manifestation of microscopic motion. Computing the classical resonant frequencies of atoms or molecules in a lattice gives numbers of the order of $10^{13}$ Hz, that is in the infrared, so when molecules jiggle they give off heat. These lattice vibrations are called phonons. Phonons have both wave-like and particle-like aspects. Lattice vibrations are responsible for the transport of heat, and we know that is a diffusive phenomenon, described by the Fourier equation. However, if the lattice is cooled to near absolute zero, the mean free scattering of the phonons becomes comparable to the macroscopic size of the sample. When this happens, lattice vibrations no longer behave diffusively but are actually wave-like or thermal wave. By controlling the temperature of a sample, one can control the extent to which heat is ballistic (thermal wave) or diffusive. In essence if a heat pulse is launched into sample (by the laser pulse interaction) and if the phonons can get across the sample without scattering, they will propagates as thermal waves.

The extent to which the motion of quasiparticles (phonons) or particles is ballistic, is described by the value of the relaxation time, $\tau$. For ballistic (wave-like) motion, $\tau \to \infty$. The equation which is generalization of the Fourier equation (in which $\tau \to \infty$) is the Heaviside equation [6] for thermal processes:

$$\tau \frac{\partial^2 T}{\partial t^2} + \frac{\partial T}{\partial t} = D \nabla^2 T \qquad (1)$$

For very short relaxation time, $\tau \to 0$ we obtain from equation (1) the Fourier equation

$$\frac{\partial T}{\partial t} = D \nabla^2 T \qquad (2)$$

and for $\tau \to \infty$ we obtain from formula (1), the ballistic $\equiv$ thermal wave motion:

$$\frac{1}{v^2} \frac{\partial^2 T}{\partial t^2} = \nabla^2 T. \qquad (3)$$

In the set of papers [6,7,8] the quantum generalization of the Heaviside equation was obtained and solved:

$$\frac{1}{v^2} \frac{\partial^2 T}{\partial t^2} + \frac{m}{\hbar} \frac{\partial T}{\partial t} = \frac{\partial^2 T}{\partial t^2}, \qquad (4)$$

where $v = \alpha c$ and is the fine structure constant, $c$ is the vacuum light velocity. In formula (4) $m$ is the *heaton* mass [3]. *Heaton* energy is equal

$$E_h = m\alpha^2 c^2. \tag{5}$$

In papers [6] Heaviside equation was obtained for the fermionic gases (electrons, nucleons, quarks). In this paper the Heaviside equation will be obtained for particles with mass *m*, where *m* is the mass of the fermion or boson. Moreover beside the elastic scattering of the particles, the creation and absorption of the heat carriers will be discussed. The new form of the discrete Heaviside equation will obtained as the result of the discretization of the one-dimensional Boltzmann equation. The solution of the discrete Boltzmann equation will be obtained for Cauchy boundary conditions, initialed by ultra-short laser pulses, i.e. for $\Delta t \leq \tau$, the relaxation time.

## 2. The model

Let us consider the one-dimensional rod (strand) which can transport "particles" – heat carriers. These particles, however may move only to the right or to the left on the rod. Moving particles may collide with the fixed scatter centra, barriers, dislocations) the probabilities of such collisions and their expected results being specified. All particles will be of the same kind, with the same energy and other physical specifications distinguishable only by their direction.

Let us define:

$u(z,t)$ = expected density of particles at *z* and at time *t* moving to the right,

$v(z,t)$ = expected density of particles at *z* and at time *t* moving to the left.

Furthermore, let

$\delta(z)$ = probability of collision occurring between a fixed scattering centrum and a particle moving between *z* and $z + \Delta$.

Suppose that a collision might result in the disappearance of the moving particle without new particle appearing. Such a phenomenon is called *absorption*. Or the moving particle may be reversed in direction or back-scattered. We shall agreeing that in each collision at *z* an expected total of $F(z)$ particles arises moving in the direction of the original particle, $B(z)$ arise going in the opposite direction.

The expected total number of right-moving particles in $z_1 \leq z \leq z_2$ at time *t* is

$$\int_{z_1}^{z_2} u(z,t) dz, \tag{6}$$

while the total number of particles passing *z* to the right in the time interval $t_1 \leq t \leq t_2$ is

$$w \int_{t_1}^{t_2} u(z,t) dt, \tag{7}$$

where *w* is the particles speed.

Consider the particle moving to the right and passing $z + \Delta$ in the time interval $t_1 + \Delta/w \leq t \leq t_2 + \Delta/w$:

$$w \int_{t_1 + \Delta/w}^{t_2 + \Delta/w} u(z+\Delta, t') dt' = w \int_{t_1}^{t_2} u\left(z + \Delta, t' + \frac{\Delta}{w}\right) dt'. \tag{8}$$

These can arise from particles which passed z in the time interval $t_1 \leq t \leq t_2$ and came through ($z, z+\Delta$) without collision

$$w \int_{t_1}^{t_2} (1 - \Delta \delta(z,t')) u(z,t') dt' \tag{9}$$

plus contributions from collisions in the interval $(z, z+\Delta)$. The right-moving particles interacting in $(z, z+\Delta)$ produce in the time $t_1$ to $t_2$,

$$w \int_{t_1}^{t_2} \Delta \delta(z,t') F(z,t') u(z,t') dt' \tag{10}$$

particles to the right, while the left moving ones give:

$$w \int_{t_1}^{t_2} \Delta \delta(z,t') B(z,t') v(z,t') dt' . \tag{11}$$

Thus

$$w \int_{t_1}^{t_2} u\left(z+\Delta, t'+\frac{\Delta}{w}\right) dt' = w \int_{t_1}^{t_2} u(z,t') dt' + w\Delta \int_{t_1}^{t_2} \delta(z,t')(F(z,t')-1) u(z,t') dt'$$
$$+ w\Delta \int_{t_1}^{t_2} \delta(z,t') B(z,t') v(z,t') dt'. \tag{12}$$

Now, we can write:

$$u\left(z+\Delta, t'+\frac{\Delta}{w}\right) = u(z,t') + \left(\frac{\partial u}{\partial z}(z,t') + \frac{1}{w}\frac{\partial u}{\partial t}(z,t')\right)\Delta \tag{13}$$

to get

$$\int_{t_1}^{t_2} \left(\frac{\partial u}{\partial z}(z,t')\right) + \frac{1}{w}\frac{\partial u}{\partial t}(z,t') dt' = \int_{t_1}^{t_2} \delta(z,t')((F(z,t')-1)u(z,t') + B(z,t')v(z,t'))dt'. \tag{14}$$

On letting $\Delta \to 0$ and differentiating with respect to $t_2$ we find

$$\frac{\partial u}{\partial z} + \frac{1}{w}\frac{\partial u}{\partial t} = \delta(z,t)(F(z,t)-1)u(z,t) + \delta(z,t)B(z,t)v(z,t). \tag{15}$$

In a like manner

$$-\frac{\partial v}{\partial z} + \frac{1}{w}\frac{\partial v}{\partial t} = \delta(z,t)B(z,t)u(z,t) + \delta(z,t)(F(z,t)-1)v(z,t). \tag{16}$$

The system of partial differential equations of hyperbolic type (15,16) is the Boltzmann equation for one dimensional transport phenomena [11].
Let us define the total density for heat carriers, $\rho(z,t)$

$$\rho(z,t) = u(z,t) + v(z,t) \tag{17}$$

and density of heat current

$$j(z,t) = w(u(z,t) - v(z,t)). \tag{18}$$

Considering equations (15–18) one obtains

$$\frac{\partial \rho}{\partial z} + \frac{1}{w^2}\frac{\partial j}{\partial t} = \delta(z,t)u(z,t)(F(z,t)-B(z,t)-1) + \delta(z,t)v(z,t)(B(z,t)-F(z,t)+1). \tag{19}$$

Equation (19) can be written as

$$\frac{\partial \rho}{\partial z} + \frac{1}{w^2}\frac{\partial j}{\partial t} = \frac{\delta(z,t)(F(z,t)-B(z,t)-1)j}{w} \tag{20}$$

or

$$j = \frac{w}{\delta(z,t)(F(z,t)-B(z,t)-1)}\frac{\partial \rho}{\partial z} + \frac{1}{w\delta(z,t)(F(z,t)-B(z,t)-1)}\frac{\partial j}{\partial t}. \tag{21}$$

Denoting, $D$, diffusion coefficient

$$D = -\frac{w}{\delta(z,t)(F(z,t)-B(z,t)-1)}$$

and $\tau$, relaxation time

$$\tau = \frac{1}{w\delta(z,t)(1-F(z,t)-B(z,t))} \qquad (22)$$

equation (21) takes the form

$$j = -D\frac{\partial \rho}{\partial z} - \tau\frac{\partial j}{\partial t}. \qquad (23)$$

Equation (23) is the Cattaneo's type equation and is the generalization of the Fourier equation. Now in a like manner we obtain from equation (15–18)

$$\frac{1}{w}\frac{\partial j}{\partial z} + \frac{1}{w}\frac{\partial \rho}{\partial t} = \delta(z,t)u(z,t)(F(z,t)-1+B(z,t))$$
$$+\delta(z,t)v(z,t)(B(z,t)+F(z,t)-1)) \qquad (24)$$

or

$$\frac{\partial j}{\partial z} + \frac{\partial \rho}{\partial t} = 0. \qquad (25)$$

Equation (25) describes the conservation of energy in the transport processes.
Considering equations (23) and (25) for the constant $D$ and $\tau$ the hyperbolic Heaviside equation is obtained:

$$\tau\frac{\partial^2 \rho}{\partial t^2} + \frac{\partial \rho}{\partial t} = D\frac{\partial^2 \rho}{\partial z^2}. \qquad (26)$$

In the case of the *heaton* gas with temperature $T(z,t)$ equation (26) has the form [6,7,8]

$$\tau\frac{\partial^2 T}{\partial t^2} + \frac{\partial T}{\partial t} = D\frac{\partial^2 T}{\partial z^2},$$

where $\tau$ is the relaxation time for the thermal processes.

## 3. The solution of the Boltzmann equation for the stationary transport phenomena in one dimensional strand

In the stationary state transport phenomena $dF(z,t)/dt = dB(z,t)dt = 0$ and $d\delta(z,t)/dt = 0$.
In that case we denote $F(z,t) = F(z) = B(z,t) = B(z) = k(z)$ and equation (10) and (11) can be written as

$$\frac{du}{dz} = \delta(z)(k-1)u(z) + \delta(z)kv(z),$$
$$-\frac{dv}{dz} = \delta(z)k(z)u(z) + \delta(z)(k(z)-1)v(z) \qquad (27)$$

with diffusion coefficient

$$D = \frac{w}{\delta(z)} \qquad (28)$$

and relaxation time

$$\tau(z) = \frac{1}{w\delta(z)(1-2k(z))}. \qquad (29)$$

The system of equations (27) can be written as

$$\frac{d^2u}{dz^2} - \frac{\frac{d}{dz}(\delta k)}{\delta k}\frac{du}{dz} + u\left[\delta^2(2k-1) + \frac{d\delta}{dz}(1-k) + \frac{\delta(k-1)}{\delta k}\frac{d(\delta k)}{dz}\right] = 0, \qquad (30)$$

$$\frac{du}{dz} = \delta(k-1)u + \delta k v(z). \qquad (31)$$

Equation (30) after difference has the form

$$\frac{d^2u}{dz^2} + f(z)\frac{du}{dz} + g(z)u(z) = 0, \qquad (32)$$

where

$$f(z) = -\frac{1}{\delta}\left(\frac{\delta}{k}\frac{dk}{dz} + \frac{d\delta}{dz}\right),$$
$$g(z) = \delta^2(z)(2k-1) - \frac{\delta}{k}\frac{dk}{dz}. \qquad (33)$$

For the constant absorption rate we put

$$k(z) = k = \text{constant} \neq \frac{1}{2}.$$

In that case

$$f(z) = -\frac{1}{\delta}\frac{d\delta}{dz},$$
$$g(z) = \delta^2(z)(zk-1). \qquad (34)$$

With functions $f(z)$ and $g(z)$ the general solution of the equation (30) has the form

$$u(z) = C_1 e^{(1-2k)^{1/2}\int \delta dz} + C_2 e^{-(1-2k)^{1/2}\int \delta dz}. \qquad (35)$$

In the subsequent we will consider the solution of the equation (32) with $f(z)$ and $g(z)$ described by (34) for Cauchy condition:

$$u(0) = q, \quad v(a) = 0. \qquad (36)$$

Boundary condition (36) describes the generation of the heat carriers (by illuminating the left end of the strand with laser pulses) with velocity $q$ heat carrier per second.

The solution has the form:

$$u(z) = \frac{2qe^{[f(0)-f(a)]}}{1+\beta e^{2[f(0)-f(a)]}}\left[\frac{(1-2k)^{\frac{1}{2}}}{(1-2k)^{\frac{1}{2}}-(k-1)}\right]\cosh[f(x)-f(a)]$$
$$+\frac{k-1}{(1-2k)^{\frac{1}{2}}-(k-1)}\sinh[f(x)-f(a)], \qquad (37)$$

$$u(z) = \frac{2qe^{(f(0)-f(a))}}{1+\beta e^{2[f(0)-f(a)]}}\left[\frac{(1-2k)^{\frac{1}{2}}+(k-1)}{k}\sinh[f(x)-f(a)]\right],$$

where

$$f(z) = (1-2k)^{\frac{1}{2}} \int \delta dz,$$

$$f(0) = (1-2k)^{\frac{1}{2}} \left[\int \delta dz\right]_0,$$

$$f(a) = (1-2k)^{\frac{1}{2}} \left[\int \delta dz\right]_a, \tag{38}$$

$$\beta = \frac{(1-2k)^{\frac{1}{2}} + (k-1)}{(1-2k)^{\frac{1}{2}} - (k-1)}.$$

Considering formulae (17), (18) and (37) we obtain for the density, $\rho(z)$ and current density $j(z)$.

$$j(z) = \frac{2qwe^{[f(0)-f(a)]}}{1+\beta e^{2[f(0)-f(a)]}} \left[\frac{(1-2k)^{\frac{1}{2}}}{(1-2k)^{\frac{1}{2}} - (k-1)} \cosh[f(z)-f(a)] - \frac{1-2k}{(1-2k)^{\frac{1}{2}} - (k-1)} \sinh[f(z)-f(a)]\right]$$

(39)

and

$$q = \frac{2qe^{[f(0)-f(a)]}}{1+\beta e^{2[f(0)-f(a)]}} \left[\frac{(1-2k)^{\frac{1}{2}}}{(1-2k)^{\frac{1}{2}} - (k-1)} \cosh[f(z)-f(a)] - \frac{1}{(1-2k)^{\frac{1}{2}} - (k-1)} \sinh[f(z)-f(a)]\right].$$

(40)

Equations (39) and (40) fulfill the generalized Fourier relation

$$j = -\frac{w}{\delta(z)} \frac{\partial \rho}{\partial z}, \qquad D = \frac{W}{\delta(z)}, \tag{41}$$

where $D$ denotes the diffusion coefficient.

Analogously we define the generalized diffusion velocity $v_D(z)$

$$v_D(z) = \frac{j(z)}{n(z)} = \frac{w(1-2k)^{\frac{1}{2}} \left[\cosh[f(z)-f(a)] - (1-2k)^{\frac{1}{2}} \sinh[f(x)-f(a)]\right]}{(1-2k)^{\frac{1}{2}} \cosh[f(x)-f(a)] - \sinh[f(x)-f(a)]}. \tag{42}$$

Assuming constant cross section for heat carriers scattering $\delta(z) = \delta_o$ we obtain from formula (38)

$$f(z) = (1-2k)^{\frac{1}{2}} z,$$
$$f(0) = 0, \tag{43}$$
$$f(a) = (1-2k)^{\frac{1}{2}} a$$

and for density $\rho(z)$ and current density $j(z)$

$$j(z) = \frac{2qwe^{-(1-2k)^{\frac{1}{2}}a\delta}}{1+\beta e^{-(1-2k)^{\frac{1}{2}}a\delta}} \left[ \frac{(1-2k)^{\frac{1}{2}}}{(1-2k)^{\frac{1}{2}} - (k-1)} \cosh\left[(2k-1)^{\frac{1}{2}}(x-a)\delta\right] \right.$$

$$\left. - \frac{(1-2k)}{(1-2k)^{\frac{1}{2}} - (k-1)} \sinh\left[(2k-1)^{\frac{1}{2}}(x-a)\delta\right] \right], \quad (44)$$

$$\rho(z) = \frac{2qe^{-(1-2k)^{\frac{1}{2}}a\delta}}{1+\beta e^{-(1-2k)^{\frac{1}{2}}a\delta}} \left[ \frac{(1-2k)^{\frac{1}{2}}}{(1-2k)^{\frac{1}{2}} - (k-1)} \cosh\left[(2k-1)^{\frac{1}{2}\delta}(x-a)\right] \right.$$

$$\left. - \frac{1}{(1-2k)^{\frac{1}{2}} - (k-1)} \sinh\left[(2k-1)^{\frac{1}{2}}(x-a)\delta\right] \right]. \quad (45)$$

We define Fourier's diffusion velocity $v_F(z)$ and diffusion length, $L$

$$v_F = \left(\frac{D}{\tau}\right)^{\frac{1}{2}}, \qquad L = v_F \tau. \quad (46)$$

Considering formulae (28) and (29) one obtains

$$v_F(z) = w(1-2k)^{\frac{1}{2}},$$

$$L = \frac{1}{\delta(1-2k)^{\frac{1}{2}}} = \frac{\lambda_{\text{mfp}}}{(1-2k)^{\frac{1}{2}}}, \quad (47)$$

where $\lambda_{\text{mfp}}$ denotes the mean free path for heat carriers.

Considering formulae (44), (45), (46), (47) one obtains

$$j(z) = \frac{2qwe^{-\frac{a}{L}}}{1+\beta e^{-\frac{a}{L}}} \left[ \frac{(1-2k)^{\frac{1}{2}}}{(1-2k)^{\frac{1}{2}} - (k-1)} \cosh\left[\frac{(x-a)}{L}\right] \right.$$

$$\left. - \frac{(1-2k)}{(1-2k)^{\frac{1}{2}} - (k-1)} \sinh\left[\frac{x-a}{L}\right] \right], \quad (48)$$

$$\rho(z) = \frac{2qe^{-\frac{a}{L}}}{1+\beta e^{-\frac{a}{L}}} \left[ \frac{(1-2k)^{\frac{1}{2}}}{(1-2k)^{\frac{1}{2}} - 1} \cosh\left[\frac{x-a}{L}\right] \right.$$

$$\left. - \frac{1}{(1-2k)^{\frac{1}{2}} - (k-1)} \sinh\left[\frac{x-a}{L}\right] \right]. \quad (49)$$

In Figs. 1–3 we present the results of the calculation for density currents of heat carriers. Figs. 1a–3a are the solution of the Fourier equation for the boundary condition (36). Figs. 1b-3b represent the solution of the Boltzmann equation formula (48) for $a \gg L, a = L$ and $a \ll L$ respectively when $k \in [0,0.5]$. For the length of strand, $a \gg L$ the both solutions Fourier and Boltzmann equations overlap. For $a \leq L$ the Boltzmann equation gives the different description of the transport processes. In that case the solution of the Boltzmann equation depends strongly on the scatterings ($k$ coefficient) of the carriers. Recently [12], the heat conduction in one-dimensional system is actively investigated. As was discussed in papers [12] the dependence of density current on $L$ can be described by the general formula

$$j \sim L^\alpha,$$

where $\alpha$ can be positive or negative. The same conclusion can be drawn from the calculation presented in our paper. In this calculation coefficient $\alpha$ depends on the scattering cross section for the heat carriers.

## 4. Conclusions

In this paper the Boltzmann equation for heat transport in one-dimensional systems is proposed and solved. It is shown that heat current density for $a \leq L$ strongly depends on the mechanism of the heat carrier scattering. The obtained results show that: (a) Boltzmann equation allows for the description of heat transport for different ratio $a/L$. (b) The heat current density strongly depends on the scattering mechanism for the heat carriers, and on the length of the strand.

**Figure captions**

Fig 1(a) The solution of the Fourier equation for the Cauchy boundary condition (formula (28)). (b) The solution of the Boltzmann equation for $a \gg L$.

Fig. 2(a) The same as in Fig. 1(a). (b) The solution of the Boltzmann equation for $a = L$.

Fig. 3(a) The same as in Fig. 1(a). (b) The solution of the Boltzmann equation for $a \ll L$.

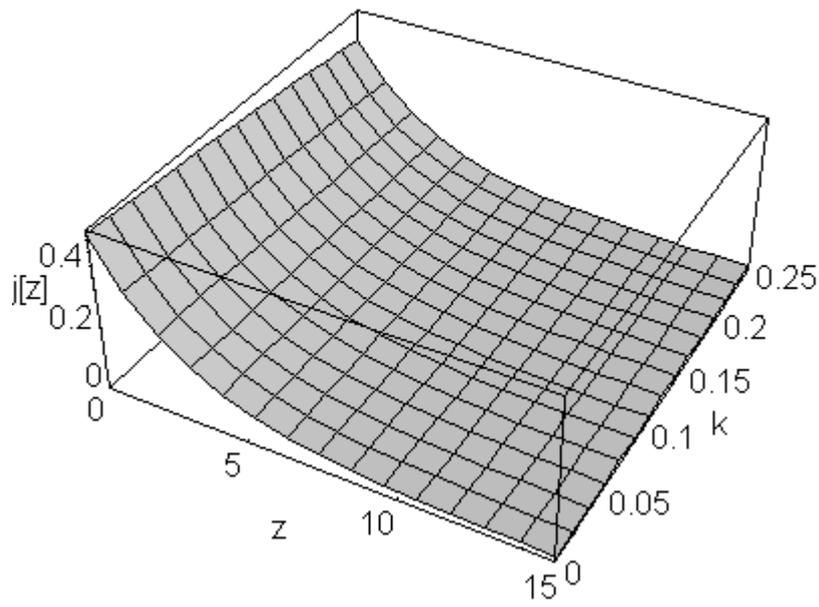

1 (a) FOURIER EQUATION

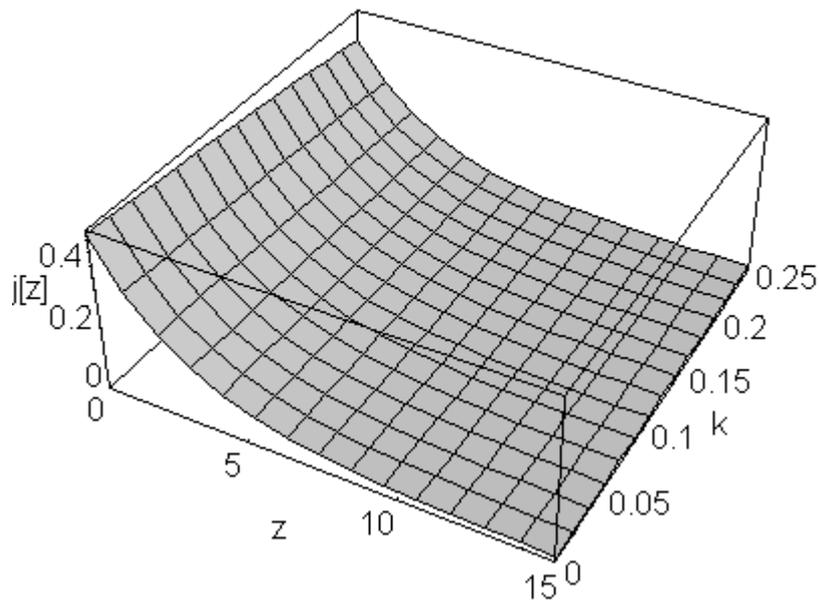

1 (b) BOLTZMANN EQUATION

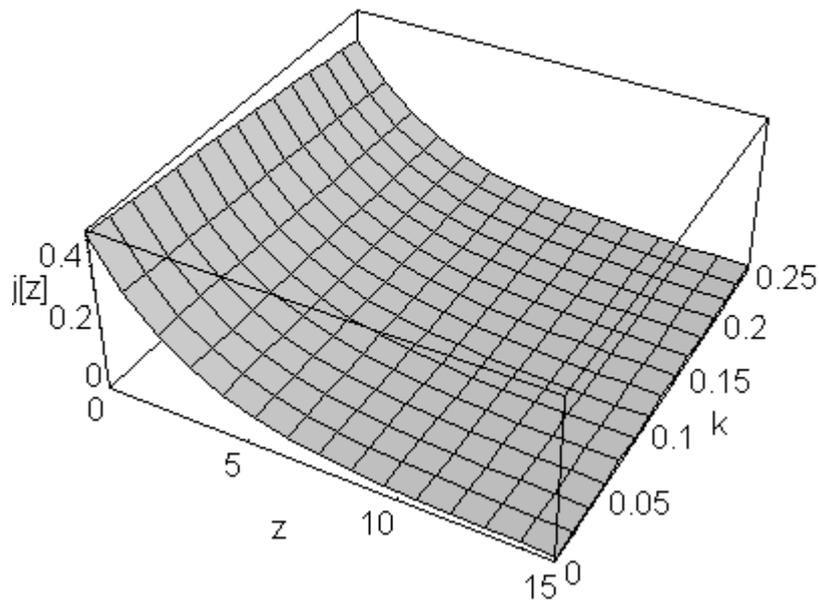

2 (a) FOURIER EQUATION

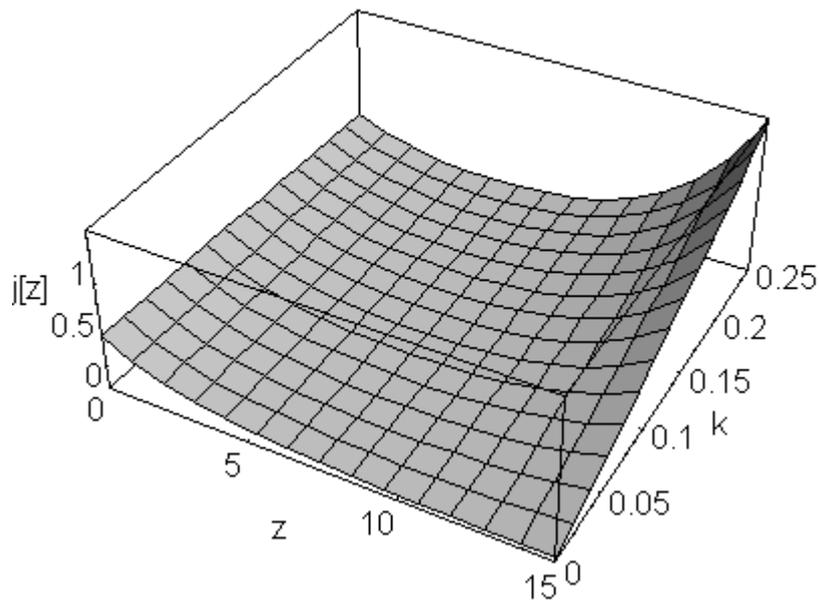

2 (b) BOLTZMANN EQUATION

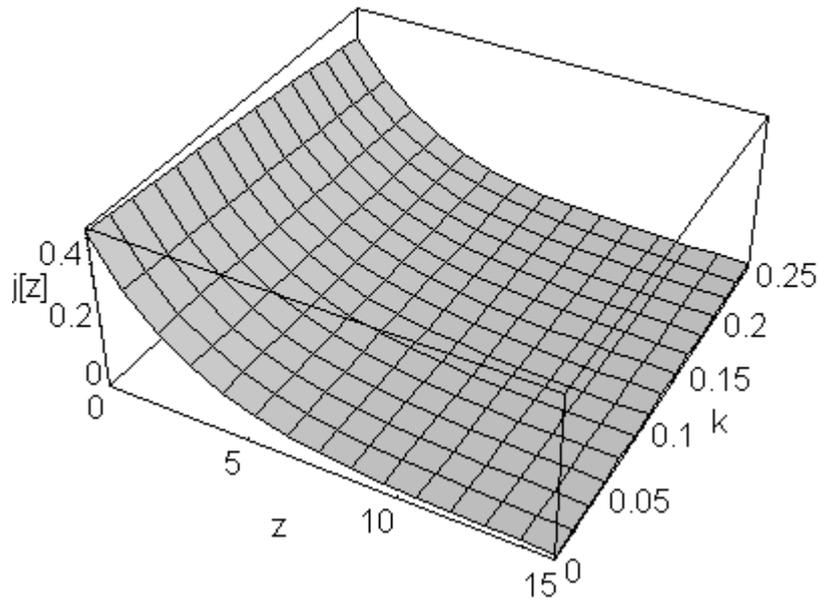

3 (a) FOURIER EQUATION

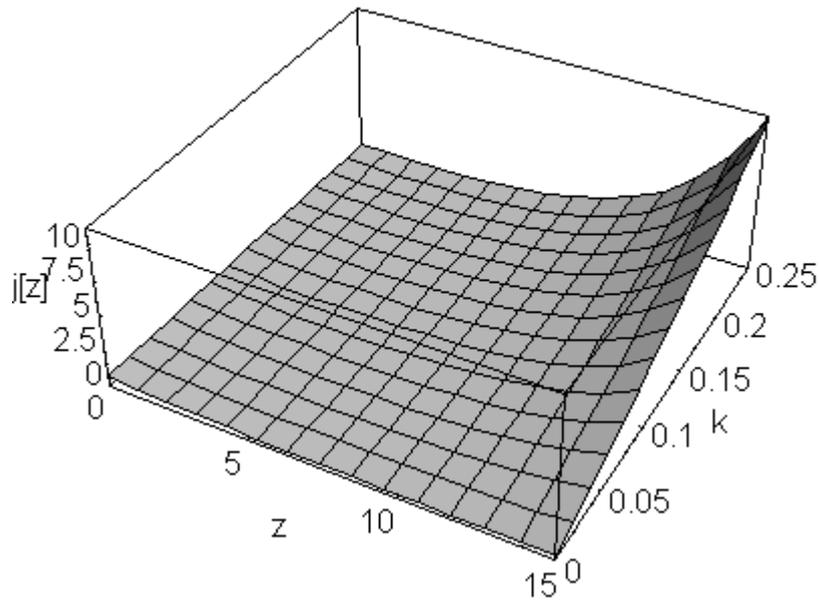

3 (b) BOLTZMANN EQUATION